# Complex High Internal-Phase Emulsions that can Form Interfacial Films with Tuneable Morphologies

Tao Li,* Ruipei Xie, Wei Chen, Andrew B. Schofield and Paul S. Clegg

Dr. T. Li, W. Chen

Wenzhou Institute, University of Chinese Academy of Sciences,

Wenzhou 325011, P. R. China.

E-mail: litao@ucas.ac.cn

Dr. T. Li, R. Xie

Beijing National Laboratory for Condensed Matter Physics and Key Laboratory of Soft Matter Physics,

Institute of Physics, Chinese Academy of Sciences,

Beijing 100190, P. R. China.

Dr. A. B. Schofield, Prof. P. S. Clegg

School of Physics and Astronomy, University of Edinburgh,

James Clerk Maxwell Building,

Peter Guthrie Tait Road, Edinburgh, EH9 3FD, UK.

*Abstract*

**High internal phase emulsions (HIPEs) are considered as an important functional material and have been the focus of intense development effort, but it has not been possible to alter their fundamental attributes at either the microcosmic or macroscopic level, which severely limits their practical applications in various areas. In this work, we report a general strategy for creating complex HIPEs and we additionally show a route to forming HIPE films at liquid interfaces. Double HIPEs and Janus HIPEs are both realized for the first time. They feature complex microscopic patterns with a short-range anisotropy, and exhibit non-Newtonian pseudoplastic flow behavior. By taking advantage of their response to a high-pH subphase, interfacial films can be successfully obtained, which are tunable in thickness and morphologies under compression.**

High internal phase emulsions (HIPEs) have a large volume fraction of dispersed phase, and therefore feature microstructures spanning from close-packed spherical droplets to distinct polyhedral geometries. Compared to their dilute counterparts, HIPEs have a very large interfacial area, exhibit unique viscoelastic behavior and can be widely used as templates for key functional materials.[1] Conventional HIPEs are a class of three-component soft matter with one continuous phase, one dispersed phase, and necessary stabilizers.[2] By learning from developments with in other ternary systems, researchers on HIPEs have made big steps forwards in recent years. For instance, a broad range of colloidal particles have been applied to stabilize HIPEs (*i.e.*, Pickering-HIPEs). [1,3-5] Non-aqueous HIPEs have been created to carry out water-sensitive or high temperature reactions.[6] Furthermore, besides homogenous emulsification, sonochemical synthesis has been demonstrated to be an alternative approach for fabricating HIPEs.[7]

These achievements, although they cover almost all aspects of HIPEs, have not altered their fundamental attributes, neither the microscopic pattern nor the macroscopic configuration. Correspondingly, the inherent limitations of HIPEs also remain (e.g., they do not have the capacity to hold chemically distinct materials).[8] One area of progress is the formation of emulsified systems with multiple-component dispersed phases, whose internal microstructure can feature separated droplets,[9] core-shell droplets,[10-12] or Janus droplets.[9,10] Droplet geometry and composition have a crucial influence on the emulsion properties and functionality, making the complex emulsions increasingly important in various application fields.[10] However, complex HIPEs with anisotropic internal microstructure have not yet been achieved. On the other hand, the solid-like flow behavior caused by the close packing of droplets makes HIPEs promising candidates for two dimensional (2D) soft materials,[13] but such implementations are also limited so far.

In this work, by including the second dispersed phase, we report for the first time complex HIPEs based on water, a pair of immiscible oils, nanoparticles and an anionic surfactant, sodium dodecyl sulfate (SDS). Both double HIPEs and Janus HIPEs with anisotropic configurations and unique viscoelastic properties can be realized. By drop-casting the complex HIPE onto the surface of a high-pH subphase, interfacial films of droplets can be obtained. We demonstrate that the films can be one-droplet thick, and exhibit a gas-liquid phase transition when being compressed in a Langmuir trough.

Figure 1a demonstrates the micro-structure of the created system when shearing 1.5g of silicone oil, 1.5g of castor oil and 0.5g of aqueous solution (water with 4mg of SDS and 6mg of silica nanoparticles) at 8000 rpm (~13700 s$^{-1}$) for 40 seconds. Spherical droplets of silicone oil (black) and castor oil (red) are closely packed as can be seen. The two different types of droplets, while fairly comparable in size (average diameter of 6.5±0.8 μm) and shape, appear to be mutually independent, *i.e.*, this is a double HIPE system. Similar to conventional HIPEs, the continuous phase (green) forms a 3D network to stabilize the entire system; capillary water bridges bind and separate individual droplets (Figure 1b). Although the distribution of droplets formed by different oils is irregular in general, some random arrays of one type of droplets can still be observed within localized regions (Figure 1c). When different types of droplets are not distinguished, analysis performed by 2D fast Fourier transform (2D-FFT) indicates a ~6.8 μm structure periodicity (Figure 1d),[14] coinciding with the droplet diameter. This periodicity disappears when only analyzing the castor oil droplets in the same region. The corresponding 2D-FFT pattern (Figure 1e) illustrates that, the spatial frequency has slightly higher intensity in both horizontal and vertical directions, reflecting a local orientation of the droplet arrays. This becomes more obvious when the arrays exhibit some ordering in the vertical or horizontal direction (Figure S1, Supporting Information).[15] Together, these confirm a short-range anisotropy of double HIPE's micro-structure.

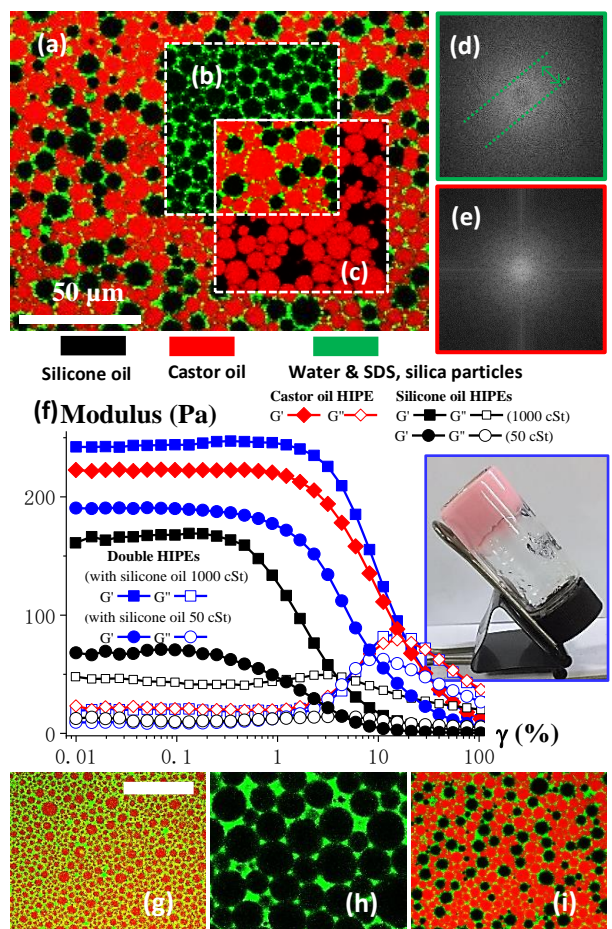

**Figure 1.** a-c) Confocal micrographs of the created double HIPE (a), the FITC labeled continuous network (b) and the Nile red labeled castor oil droplets (c). d), e) 2D-FFT patterns of the continuous network (d, the green dashed lines and the arrow mark a ~ 6.8 μm structure periodicity) and the random arrays of castor oil droplets in the same region (e). f) The measured G' and G'' of the double HIPE (blue), two silicone oil-HIPEs (50cSt & 1000 cSt, black) and one castor oil-HIPE (red). Inset illustrates the macroscopic configuration of the double HIPE. g-i) The micro-structures of the castor oil-HIPE, silicone oil-HIPE (50 cSt) and double HIPE, respectively. The scale bar is 50 μm.

Figure 1f compares the linear viscoelasticity of two double HIPEs (blue), two "single" HIPEs formed just by silicone oil (black), and one HIPE of castor oil (red). All systems exhibit a general signature of yield stress materials;[1] the elastic modulus (G') and loss modulus (G'') are both linear for a large amplitude of strain. Clearly, the G' of silicone oil-HIPEs strongly depends on the oil's viscosity[12,16] and the G' of castor oil-HIPE is larger than both silicone oil-HIPEs. When using castor oil (viscosity ~600 cSt) and a less viscous silicone oil (50 cSt), the G' of the created double HIPE falls in between (blue circle). But with a very viscous silicone oil (1000 cSt), the double HIPE exhibits a larger G' than both "single" HIPEs (blue square). It indicates that, the two internal phases may generate a synergistic effect, making the combined system more than a simple summation of two "single" HIPEs. This is directly embodied in the radius $d$ of the droplets. As shown in Figure 1g, $d$ of the castor oil-HIPE is 2.7 ± 0.2 μm, while all HIPEs formed by different viscous silicone oils have a 3-4 times larger $d$ (Figure 1h and Figure S2). Mixing these two liquids remarkably decreases the size of the silicone oil droplets, making them comparable to the castor oil ones (Figure 1i and Figure S2). Meanwhile, this synergistic effect may get influenced by the viscosity ratio between castor oil and silicone oil, which has been proved in previous studies of ternary polymer blends,[17,18] although the related mechanisms are still not clear (see more discussions in Supporting Information).

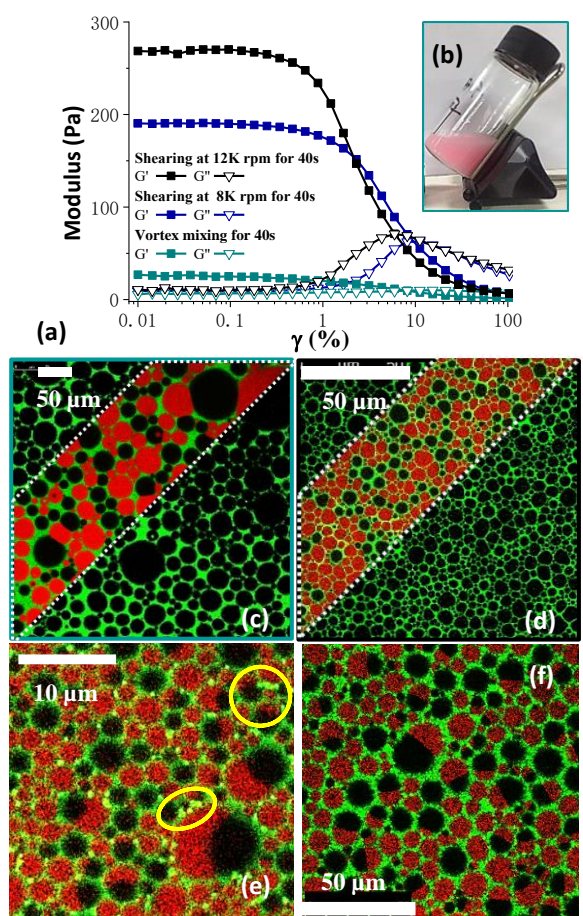

*Figure 2*. a) The measured G' of double HIPEs increases significantly with the shearing rate. b) A sample formed by vortex mixing remains partially fluidic. c), d) Microstructures of a fluidic system (c) and a highly gelled system (d). The fluorescent signal from the Nile red is shown only inside the dashed box. e), f) Janus HIPEs created by adjusting the interfacial tensions between the liquids. Some particles aggregate around the droplets (e, marked in yellow) when adding a modest amount of ethanol into the aqueous phase.

When using a certain viscous silicone oil, the viscoelastic properties of double HIPEs strongly depend on the mechanical energy employed during system fabrication. Its G' increases significantly with the shearing rate $\dot{\gamma}$ (Figure 2a); systems formed by vortex mixing ($\dot{\gamma}\sim5100$ s$^{-1}$) remain partially fluidic (Figure 2b). Figure 2c and d illustrate the microstructures of a fluidic system and a highly gelled system ($\dot{\gamma}\sim$12K rpm or $\sim20550$ s$^{-1}$) respectively. For contrast, the images mainly show the network formed by the continuous phase. Clearly, in Figure 2c, the droplets are large ($d \sim 13.4 \pm 7.2$ µm), spherical, and have a large distance between each other (> 4µm). By contrast, Figure 2d demonstrates a highly concentrated emulsion of small polyhedral droplets ($d \sim 5.9 \pm 1.7$ µm), which can be recognized as a typical double HIPE.

Similar to most emulsion-based materials, interfacial tension $\sigma$ plays an important role in determining the morphology of our systems. Adding a modest amount of ethanol into the aqueous phase can increase $\sigma$ between water and the oils (Table S1, Supporting Information). As a result, the created HIPE exhibits Janus topology (Figure 2e). Ethanol can also modify the surface chemistry of the added nanoparticles, making them easier to adsorb onto the oil-water interface. In this case, the interface can be more efficiently stabilized, and the created Janus droplets are hence smaller. Some aggregated particles (highlighted in Figure 2e) can be found around the droplets, causing a heterogeneous network. Janus morphology can also be achieved by using different oil pairs. Figure 2f demonstrates a Janus HIPE with sunflower oil (red) replacing the castor oil. Here, the Janus topology should be caused by the decrease of $\sigma$ between two oils (Table S2, Supporting Information).

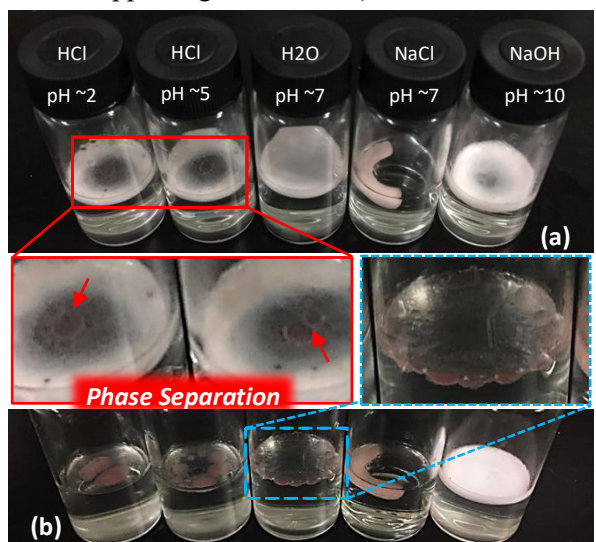

*Figure 3.* The phase behavior of a pipetted double HIPE at an air-water interface with different pH values upon initial spreading (a) and after two weeks (b). The red and blue rectangles mark a rapid and a slow complete phase separation respectively.

The complex HIPEs are stabilized by negative charge at the interface and a nanoparticle-occupied viscoelastic network in the continuous phase, it is therefore possible to create an interfacial film of these droplets by taking advantages of their response to the pH of a subphase.[19] To test this, an air-water interface with different pH values was prepared and a double HIPE was pipetted onto it. Rapid phase separation occurs when lowering the water pH with HCl

(marked in the red rectangle, Figure 3a); slow but complete phase separation can also be observed on pure water surface within 14 days (marked in the blue rectangle, Figure 3b). Intriguingly, the pipetted sample keeps its bulk shape at the surface of water when 1M NaCl is added, neither spreading at the interface, nor dispersing into the subphase (see the fourth sample, Figure 3a and b). Further analysis is needed to confirm the mechanism behind this phenomenon, which may involve the precipitation of SDS,[20] or the interaction between $Na^+$ and the particles.[21] A film of droplets is truly realized at the surface of a NaOH solution (~10 mM, pH ~10), where the droplets stay and spread uniformly (see the fifth sample).

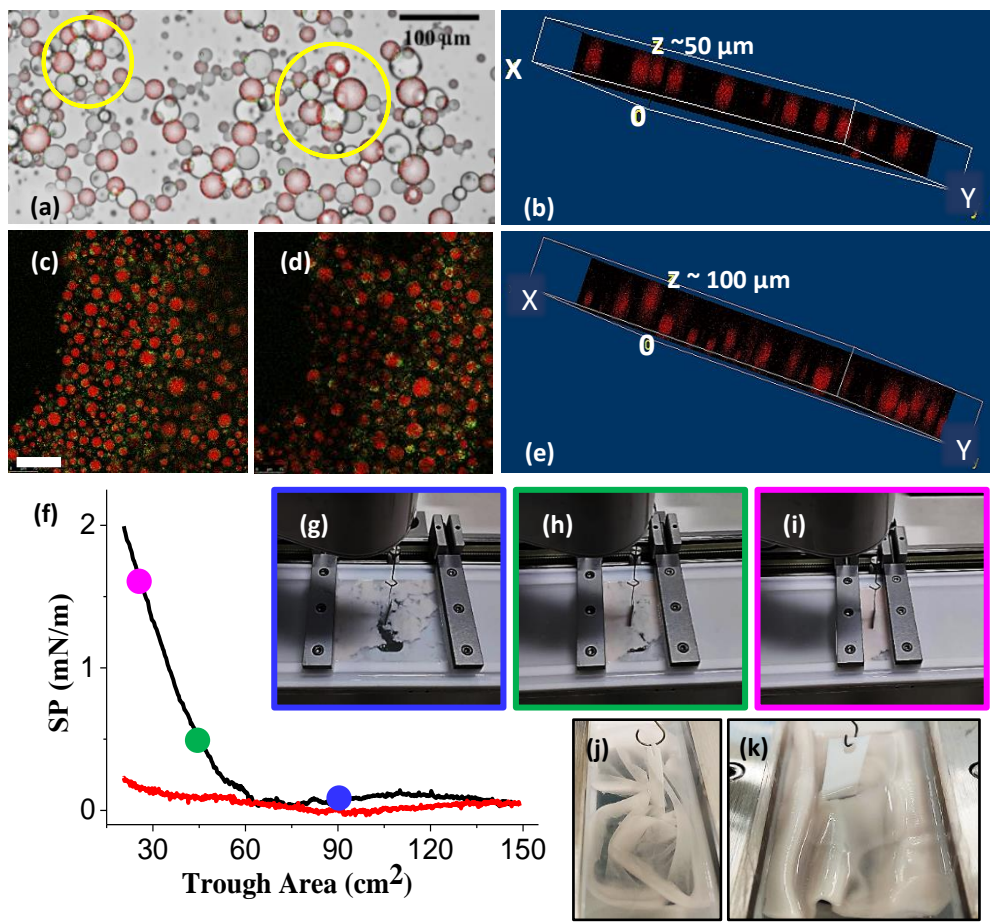

*Figure 4.* a), b) A monolayer-like film of binary droplets (a) and the corresponding 3D scanning (b). Some droplet aggregations are marked in yellow circles. c-e) A interfacial bilayer of oil droplets that can be focused on two different planes (c,d) (scale bar = 75 μm) and its 3D scanning (e). f) The Π-A isotherm of an empty interface (the red curve) and that of a film containing binary droplets (the black curve). g-i) The interface at certain compression stages as marked in (f). j), k) Gelled interfacial films with different thickness were compressed in a Langmuir trough.

Figure 4a (taken directly from the water surface) demonstrates that the trapped droplets retain their original size, shape, and the binary composition (red droplets are castor oil droplets, the rest are silicone oil droplets); no obvious break-up or coalescence was observed. Meanwhile, the network formed by the continuous phase disappears, making a very large area per droplet. However, the droplets do not distribute separately, but connect with each other into arrays and random aggregations (marked in yellow). 3D scanning confirms that it is a monolayer-like film with most droplets being in the same plane (Figure 4b). By adjusting the amount of the spreading sample, or the available area of the interface, multilayers with more complex morphologies can be produced. Figure 4c and d illustrate the fluorescent castor oil droplets focused on two different planes, which suggest a bilayer structure (see also its 3D scanning, Figure 4e).

By compressing the interfacial sheets in a Langmuir trough, a surface pressure-area (Π-A) isotherm is obtained (the black curve, Figure 4f). Similar to the phase behaviours of a particle monolayer, the interfacial droplets undergo a phase transition from a gas phase (Figure 4g) to a liquid phase (Figure 4h). Further compression leads to a more dense film at the interface (Figure 4i), and a significant increase of the measured Π. However, this increase cannot reach a high value, and the compressed film does not collapse like a jammed monolayer of hard-sphere colloids.[22] This can be related to the precipitation of SDS in a high pH environment, which makes the interaction between the droplets very weak. As a result, the droplets can easily re-arrange themselves and from a multiple layer, similar to a 2D wet foam.[23,24] If adding sodium alginate into the aqueous phase and using $Ca^{2+}$ for ionic cross-linking, the created films exhibit more significant material properties. Thin films (~ 30μm) can be soft and smooth as silk (see Figure 4j); the thicker ones are highly elastic and form distinct wrinkles under compression (Figure 4k). Moreover, a thixotropy study on the double HIPEs has demonstrated a hysteresis loop (Figure S6), which indicates a partial structure recovery and is common in topical transdermal drug delivery systems.[12,18,25] Therefore, using these films, we will be able to develop a new type of medical skin patch that can deliver from three different phases with various active ingredients.

With parallel but separate functionalities provided by multiple phases, the created complex HIPEs can greatly expand the applications of liquid materials in food, cosmetics, and medical engineering. The obtained interfacial films of droplets also represent an important step towards producing 2D soft materials with unique functionality. Such films will be able to transport/deliver components that are separately soluble in variously water, a vegetable oil and a silicone oil. The heterogenous environments could be employed in composite fabrication - or in delivery applications where the chemical identity of the channels is maintained.

*Conflicts of interest*

There are no conflicts to declare.

*Acknowledgements*

We thank Professor Jure Dobnikar, Fangfu Ye and Ke Chen from the Institute of Physics, Chinese Academy of Sciences, Professor Ryohei Seto from Wenzhou Institute, University of Chinese Academy of Sciences (WIUCAS). This work was supported by the Young Scientists Fund of the National Natural Science Foundation of China (Grant No. 11904390) and the Scientific Research Starting Foundation from WIUCAS (WIUCASQD2020003).

# Supporting Information

**Experimental Section**

**Materials.** Silicone oil (10 cSt, 50 cSt, and 1000 cSt), Fluorescein isothiocyanate (FITC)–dextran (average molecular weight 3,000-5,000, FD-4) and Sodium dodecyl sulfate (SDS) were purchased from Sigma Aldrich and used as received. Sunflower oil (chemical grade, batch number: BCBV5320) and castor oil were purchased from Sigma Aldrich and dyed with Nile Red (Sigma). Deionized water was obtained from a Milli-Q water purification system (Millipore, Bedford, MA).

**Emulsification.** 0.5g of aqueous phase containing a low concentration ~6 mg of silica nanoparticles (~150 nm in diameter, synthesized in the laboratory and fluorescently labelled with FITC) and ~4 mg of SDS was first prepared. Then 1.5g of silicone oil was added to the aqueous solution, followed by 1.5g of castor oil. The entire blend was first stirred by vortex mixing (at ~5100 $s^{-1}$) for ~20s, and then sheared using an Ultra-Turrax homogenizer (IKA T18 basic) with a 10 mm diameter head operating at ~13700 $s^{-1}$ (8000 rpm) for 40s.

**Characterization.** The imaging was performed using a confocal microscope (Leica, SP8, Germany) with a 10× objective (N.A. =1.4). Fluorescence excitation was provided by a 488 nm laser (for FITC) and a 555 nm laser (for Nile Red); emission filters were used as appropriate. Interfacial tensions were measured using a pendent drop technique with an OCA-20 contact angle system (Dataphysics, Germany). The Young's modulus of individual droplets was measured with a Piuma Nanoindenter (Optics11, BV, the Netherlands). All pH values were measured using a FiveEasy Plus pH meter (Mettler-Toledo Instruments, Shanghai, China). The rheological experiments were performed on a TA Instruments AR 2000 rheometer (USA). A thixotropy study was carried out with an RST Rheometer (Brookefiled, USA). Surface pressure measurements were performed in a Langmuir trough (KSV) with symmetric barriers. The barriers were typically moved together at a rate of 5 mm/min.

**Figure S1, Two-dimensional fast Fourier transform (2D-FFT) patterns of castor oil droplets**. Generally, when the droplets were aligned in one direction, the spatial frequency parallel with the direction is low and the spatial frequency perpendicular to the direction is high. The spatial frequency in pattern (a) exhibits a slightly higher intensity in both horizontal and vertical directions, reflecting a random distribution of the droplets. Pattern (b) has a higher intensity of the spatial frequency in horizontal direction, suggesting that the droplets arrays are relative ordering in the vertical direction. Pattern (c) indicates that the droplets are more ordering in both directions.

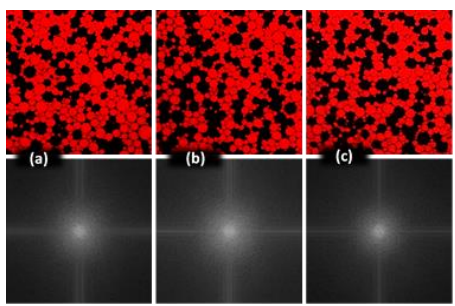

**Figure S2, The micro-structures of "single" HIPEs and double HIPEs (the scale bar =50 µm).** All systems are created by shearing at 8K rpm for 40s; the water phase contains SDS and silica nanoparticles.

(a) A HIPE system formed just by castor oil (~600 cSt). Droplet diameter = 5.4±0.4 µm; average distance between droplets ~ 1.1 µm.

(b) A "single" HIPE formed just by silicone oil (10 cSt). Droplet diameter = 21.6±0.8 µm; average distance between droplets ~ 3.8 µm. (c) A double HIPE formed by silicone oil (10 cSt) and castor oil. Droplet diameter = 6.5±0.8 µm; average distance between droplets ~ 1.3 µm.

(d) A "single" HIPE formed just by silicone oil (50 cSt). Droplet diameter = 22.1±0.7 µm; average distance between droplets ~ 3.1 µm. (e) A double HIPE formed by silicone oil (50 cSt) and castor oil. Droplet diameter =6.4±0.8 µm; average distance between droplets ~ 1.2 µm.

(f) A "single" HIPE formed just by silicone oil (1000 cSt). Droplet diameter = 17.6±0.7 µm; average distance between droplets ~ 3.4 µm. (g) A double HIPE formed by silicone oil (1000 cSt) and castor oil. Droplet diameter =6.7±1.1 µm; average distance between droplets ~ 0.9 µm.

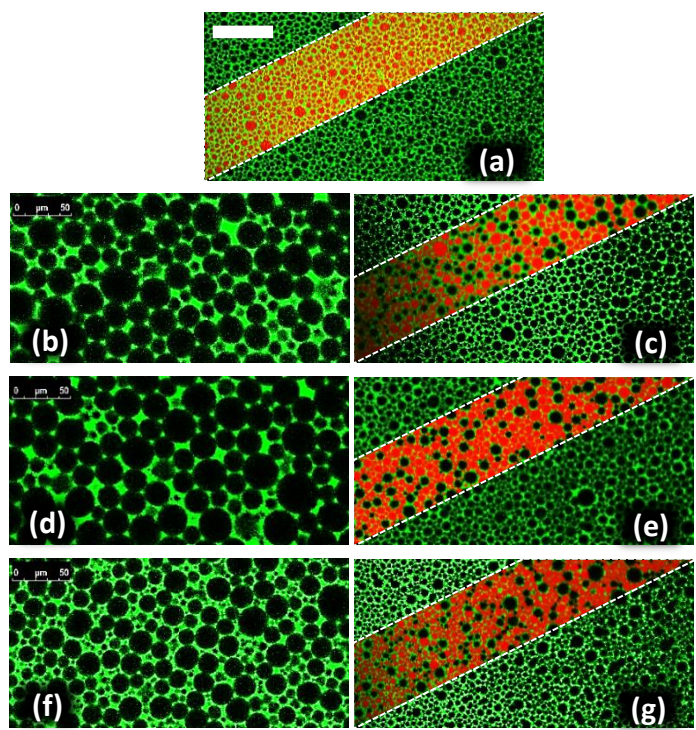

**Figure S3,** Thixotropic profile of the double HIPE measured versus shear rate at room temperature.

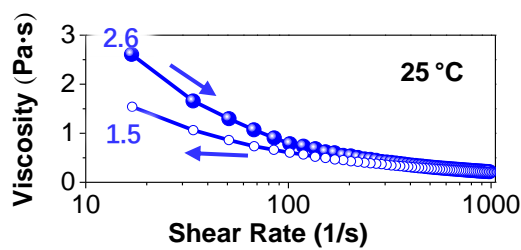

**Table S1,** The measured interfacial tensions between water, silicone oil and castor oil with the added fluorescent dyes and surfactants. All the values are obtained statically at room temperature. SDS adsorbs at the oil–water interface rapidly, which therefore significantly reduces the interfacial tensions between water and the oil.

| Liquid & liquid | Measured interfacial tension |
|---|---|
| Silicone oil (50 mPa·s) & Water | ~ 37.2 mN/m |
| Silicone oil (50 mPa·s) & water (with SDS) / Water (with SDS and Ethanol) | ~12.0 mN/m /~13.3 mN/m |
| Castor oil (with Nile red) & Water | ~ 23.1 mN/m |
| Castor oil (with Nile red) & Water (with SDS) / Water (with SDS and Ethanol) | ~ 2.2 mN/m /~5.8 mN/m |
| Silicone oil (50 mPa·s) & Castor oil (with Nile red) | ~ 3.2 mN/m |

**Table S2,** The measured interfacial tensions between water, silicone oil and sunflower oil with the added fluorescent dyes and surfactants. All the values are obtained statically at room temperature.

| Liquid & liquid | Measured interfacial tension |
|---|---|
| Silicone oil (50 mPa·s) & Water | ~ 37.2 mN/m |
| Silicone oil (50 mPa·s) & Water (with SDS) | ~12.0 mN/m |
| Sunflower oil (with Nile red) & Water | ~ 23.1 mN/m |
| Sunflower oil (with Nile red) & Water (with SDS) | ~3.2 mN/m |
| Silicone oil (50 mPa·s) & Sunflower oil (with Nile red) | ~ 2.1 mN/m |

**Discussion on Princen and Kiss model**

In 1986, Princen and Kiss (see [*J. Colloid Interface Sci.* 1986, **112**, 427]) presented the following expression for static shear modulus of a series of concentrated oil-in-water emulsions, which is now widely used as an empirical model for the elastic modulus of a HIPE system:

$$G' = 1.769 \, \Phi^{\frac{1}{3}} (\Phi - 0.712) \frac{\sigma}{d} \qquad (1)$$

where $\Phi$ (> 0.712) represents the equivalent volume fraction of the dispersed phase (note $\Phi$ is not only the dispersed phase volume fraction, but also relates to the film thickness between the droplets), $\sigma$ is the interfacial tension, and $d$ is the mean radius of the droplets. The synergistic effect created between silicone oil and castor oil is first embodied in $d$. As shown in Figure 1g, $d$ of the castor oil-HIPE is 2.7 ± 0.2 μm, with an average distance ~ 1.1 μm between each other. All HIPEs formed by different viscous silicone oils have a 3-4 times larger $d$, and a larger distance (> 3μm) between the droplets (Figure 1h and Figure S2). Mixing these two liquids, obviously, decreases the size of the silicone oil droplets remarkably, making them comparable to the castor oil droplets (Figure 1i and Figure S2). Meanwhile, for the double HIPEs, the parameter $\sigma$ (which can be written as $\sigma_{double}$) is also quite complex, since it contains both $\sigma_{castor\ oil\ \&\ water}$ (~2.2 mN/m, Table S1 below) and $\sigma_{silicone\ oil\ \&\ water}$ (~12.0 mN/m, Table S1). The value of $\sigma_{double}$, however, is not necessary in between. If we apply the upper and lower bounds for the elastic properties of nanoparticle composites,[25] which is the model of springs connected in parallel or series, $\sigma_{double}$ can be smaller than $\sigma_{castor\ oil\ \&\ water}$, or larger than $\sigma_{silicone\ oil\ \&\ water}$. Currently, there is no general rule for a mixed $\sigma$. Considering the comparable size between castor oil-HIPE and the double HIPEs, we suppose the value of $\sigma_{double}$ should be somehow similar to $\sigma_{castor\ oil\ \&\ water}$. One manifestation is that, the measured G' values of both double HIPEs in Figure 1f are closer to that of the castor oil-HIPE.

Previous studies of ternary polymer blends have proved that, the viscosity ratio between two dispersed phases can strongly affect its phase morphology and rheological/mechanical properties, although the related mechanisms are still not clear. In our systems, the viscosity ratio between silicone oil and castor oil may have an influence on both $d$ and $\sigma_{double}$ via the critical capillary number ($Ca_{cri}$), which can further affect the G' of the double HIPEs.

**Discussion of the Capillary number.**

The dynamics of droplets under shear is determined by the ratio between the viscous deformation stresses (shear stress) and the restoring interfacial stresses (Laplace pressure), given by the capillary number:

$$Ca = \frac{\text{shear stress}}{\text{Laplace pressure}} = \frac{\lambda \dot{\gamma} d}{\sigma} \qquad (2)$$

where $\dot{\gamma}$ is the shear rate, $d$ is the radius of the droplets, $\sigma$ is still the interfacial tension, and $\lambda$ is the viscosity of the continuous phase. Once the value of Ca exceeds a critical threshold ($Ca_{cri}$), large droplets would deform and break up under the influence of shear. It is believed that $Ca_{cri}$ depends on the type of flow, and the viscosity ratio between the dispersed phase and the continuous phase [*Langmuir* 2011, **27**, 9760–9768; *Langmuir* 2006, **22**, 1544-1550]. The final size of the droplets is usually a balance between coalescence and breakup of the dispersed phase:

$$d \propto Ca_{cri} \frac{\sigma}{\lambda \dot{\gamma}} \qquad (3)$$

It should also be noted that, the coalescence between the droplets can be very fast and stops once the interfaces of the droplets are fully coated by stabilizers. Only droplets containing the same component can coalesce, which can be helpful to limit the droplet size in the double-HIPE systems.

We should emphasize that, the parameters involved in Equation (1)-(3) may impact on each other; the effects of some other parameters, *e.g.*, the viscosity of the dispersed phase and the liquid density, still need to be confirmed.

**Discussion of the Internal phase concentration.**

As we discussed above, $\Phi$ in Equation (1) represents the equivalent volume fraction of the dispersed phase, which relates to the finite film thickness between the droplets, *i.e.*, the distance between the droplets. Specifically,

$$\Phi^{-1/3} = \Phi_{internal}^{-1/3} - 1.105 \frac{h}{2d} \qquad (4)$$

where $h$ is the film thickness, $d$ is the droplet radius, and $\Phi_{internal}$ is the volume fraction of the internal phase.

Equation (4) is quite complex and requires some assumptions about the relationship between $h$ and $d$. If we assume that all the droplets arrange in a face-centred cubic (fcc) lattice with a distance $h$ between each other, as illustrated below:

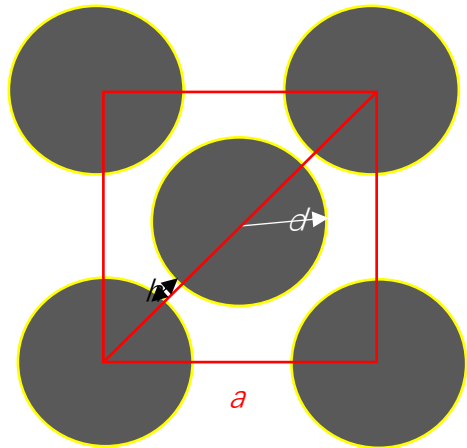

$$\sqrt{2}a = 4d + 2h \;\rightarrow\; a = \frac{4d+2h}{\sqrt{2}}$$

$$V_{cell} = a^3 = (\frac{4d+2h}{\sqrt{2}})^3$$

$$V_{droplets} = 4 \times \frac{4}{3}\pi d^3$$

the volume fraction of the dispersed phase $\Phi_{internal}$ is the same for all our systems. Therefore:

$$V_{droplets} = V_{cell} \times \Phi_{internal}$$

*i.e.*,

$$\frac{16}{3}\pi d^3 = (\frac{4d+2h}{\sqrt{2}})^3 \times \Phi_{internal} \;\rightarrow\; \frac{4\sqrt{2}}{3}\pi d^3 = (2d+h)^3 \times \Phi_{internal} \qquad (5)$$

Clearly, when $d$ gets larger, $h$ gets larger as well, consistent with our experimental observations (Figure S2).

Systems with small *d* and *h* have a high internal phase concentration, which can help to increase the G' for a HIPE system as illustrated in Equation (1), and as reported in previous experiments [Colloids and Surfaces A: Physicochem. Eng. Aspects 442 (2014) 111–122]. As shown in Figure S2b, d and f), the "single" HIPEs formed just by silicone oil have relatively low internal phase concentration, which can be part of the reason for their relatively low G'.

Shearing the system with different rate can significantly change the droplet size as suggested by Equation (2). A high shearing rate can create a large number of small droplets. During coalescence, these droplets would share the continuous phase in between, get contact with each other and lose their original spherical shape (Figure 2d). This leads to a very high internal phase concentration, and can increase the G' of the system. The opposite situation occurs when using a low shearing rate.